\documentclass[twocolumn,showpacs,preprintnumbers,amsmath,amssymb]{revtex4}
\usepackage{graphicx}
\usepackage{dcolumn}
\usepackage{bm}

\begin{document}
\preprint{APS/123-QED}

\title{Dynamics of micrometer-scale phase separation in a polymer mixture upon laser irradiation}

\author{Hirofumi Toyama}

\author{Hiroyuki Kitahata}
\author{Kenichi Yoshikawa}
\altaffiliation[To whom all the correspondence should be addressed.]{ Tel: +81-75-753-3812, Fax:+81-75-753-3779, Elecronic address: yoshikaw@scphys.kyoto-u.ac.jp}
\affiliation{Department of Physics, Graduate School of Science, Kyoto University, Kyoto 606-8502, Japan\\
}

\date{\today}

\begin{abstract}
Ten $\mu$m -sized droplets in an aqueous-two-phase system (water/ polyethylene-glycol (PEG)/ dextran) dissapear upon irradiation with a focused YAG laser. The interface of the dextran-rich droplet broadens, indicating smoothing of the concentration profile, whereas, the PEG-rich droplet shrinks and disappears. These phenomena can be described in terms of the Ginzburg-Landau free energy, by considering the laser-induced dielectric potential. 
\end{abstract}

\pacs{Valid PACS appear here}

\maketitle
It is well known that a focused laser can act as optical tweezers, and can generate attractive potential due to dielectric interaction\cite{Ashkin_1970,Ashkin,Svoboda,beads_bursting}. Optical tweezers are widely used in biology, chemistry, and physics \cite{history_laser,emulsion_laser}. Recently, it has been shown that $\mu$m-scale phase separation can be induced using a focused laser \cite{pulse_laser_droplet,micelles,association_PVCz,Mukai}. An aqueous-two-phase system consists of water mixed with two soluble polymers, and can be used as a tool to partition biomolecules. Since such systems exhibit low interfacial tension ($10^{-4}\sim 10^{-2}$ dyne/cm ) \cite{Albertsson,partitioning}, it is expected that $\mu$m-scale phase separation could easily occur upon irradiation with a focused laser. In this paper, we report the dynamics of $\mu$m-scale droplets under laser irradiation. Unexpectedly, we observed that $\mu$m-sized droplets disappear due to laser potential, and there are marked differences in the process of disappearance depending on the composition of the droplets. 

\begin{figure}
\begin{center}
  \includegraphics{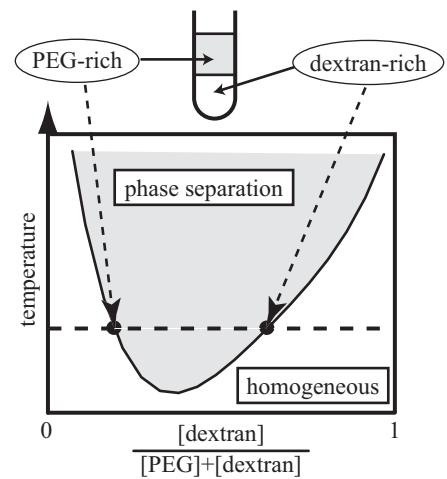}
  \caption{Schematic phase diagram of the PEG / dextran / water system. The horizontal axis is the fraction of the dextran concentration compared with the total concentration of polymers.}
  \label{fig:figure1}
\end{center}
\end{figure}

We used a system composed of water, dextran (Molecular Weight (M.W.) = 15000 $\sim$ 20000, Nacalai) and polyethylene-glycol (PEG; M.W. = 7400 $\sim$ 9000, Nacalai) as an aqueous-two-phase system. This system exhibits the phase diagram as shown in Figure \ref{fig:figure1}; phase separation is generated with an increase in temperature at a fixed composition. The phase-separated solution segregates into a PEG-rich phase (upper phase) and a dextran-rich phase (lower phase). Sample solutions were prepared by mixing water, dextran and PEG, and the mixture was allowed to stand for more than 1 hour. The composition of the sample solutions was chosen based on the phase-separating condition, just in the vicinity of the binodial line. From the phase-separated solution, a PEG-rich phase or a dextran-rich phase was transferred to a thin chamber (20 mm $\times$ 20 mm $\times$ 200 $\mu$m). We used an yttrium-aluminum-garnet (YAG) laser (wavelength = 1064 nm, Millennia IR, Spectra Physics). Since an aqueous-two-phase system has a long relaxation time, small droplets were present even after it had been allowed to stand for several days. A converged laser was passed through an oil-immersed lens (100$\times$, numerical aperture=1.3) with an inverted phase-contrast microscope (TE-300, Nikon). The experiments were carried out under constant ambient temperature at 297 $\pm$ 1 K. In the PEG-rich phase, dextran-rich droplets were monitored with a phase-contrast microscopic view, while in the dextran-rich phase, PEG-rich droplets were monitored. 
\begin{figure}
\begin{center}
  \includegraphics{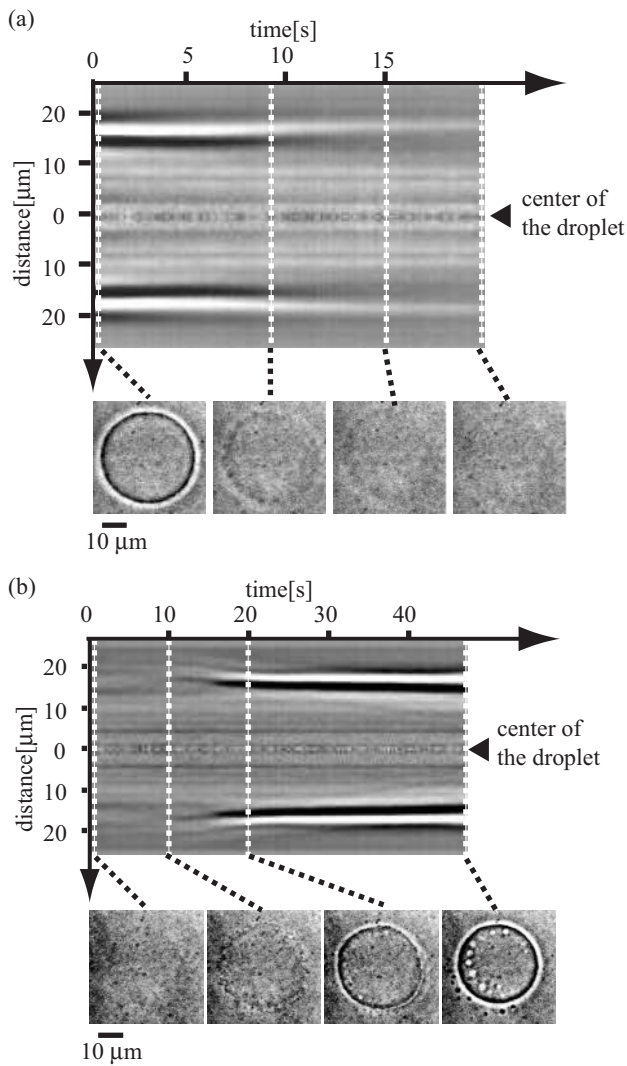}
  \caption{(a) Spatio-temporal image of a gradually disappearing dextran-rich droplet around the focus of a YAG laser ($\lambda$= 1064 nm) in the PEG-rich phase at a laser power of 1.5 W, reconstructed from time-successive video frames. Actual pictures of the droplets are shown at the bottom. (b) Spatio-temporal image of the reappearing dextran-rich droplet after stopping laser irradiation, reconstructed from successive video frames. Actual pictures of the droplets are shown at the bottom. $\mu$m-sized droplets are generated both inside and outside of the original dextran-rich droplet. The origin of the time corresponds to when the laser was switched (a) on or (b) off.}
  \label{fig:figure2}
\end{center}
\end{figure}%

Figure \ref{fig:figure2} (a) shows a spatio-temporal image of a gradually disappearing dextran-rich droplet in a PEG-rich phase at a laser power of $P$ =1.5 W, where the actual shapes of the droplets are given at the bottom. A straight line in the real pictures that passes through the center of the droplet was picked up and pasted in sequence to make the spatio-temporal image. A focused laser was used to irradiate a dextran-rich droplet in the PEG-rich phase. The interface of the droplet disappeared gradually.  When the laser was turned off, the dextran-rich droplet reappeared in the same region, followed by the formation of multiple smaller droplets both inside and outside the original dextran-rich droplet, to form a nested structure. Figure \ref{fig:figure2} (b) shows a spatio-temporal image together with the pictures of the reappearing dextran-rich droplet.  

\begin{figure}
\begin{center}
  \includegraphics{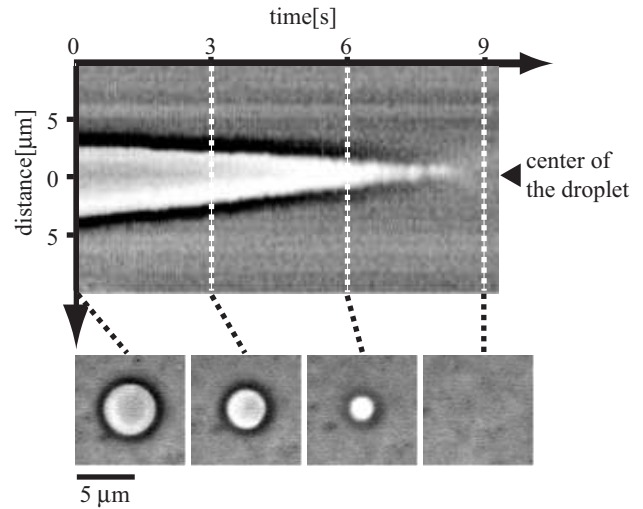}
  \caption{Spatio-temporal image of a shrinking dextran-rich droplet around the laser focus at a laser power of $P$ = 1.5 W, reconstructed from time-successive video frames. The diameter of the droplet decreased linearly with time. The origin of the time corresponds to when the laser was switched on.}
  \label{fig:figure3}
\end{center}
\end{figure}

Next, the laser was used to irradiate PEG-rich droplets in the dextran-rich phase in the same way. Figure \ref{fig:figure3} shows a spatio-temporal image and pictures of a disappearing PEG-rich droplet under laser irradiation at a laser power of $P$ =1.5 W. The diameter of the dextran-rich droplet decreased linearly with time. The PEG-rich droplet did not reappear after laser irradiation was discontinued.

The same experiments were conducted with a D$_{2}$O / dextran / PEG system, in which H$_{2}$O was replaced with D$_{2}$O. Since D$_{2}$O absorbs light with a wavelength of 1064 nm only 1/100 as much as H$_{2}$O \cite{d2o}, heating due to laser irradiation can be neglected in the experiments with D$_{2}$O. We confirmed that there was no essential difference in the experimental trends regarding the disappearance of droplets between the H$_{2}$O and D$_{2}$O solutions. Thus, the disappearing interface of droplets is attributed to the laser potential, at least as the main driving force.  Generally, when an object is smaller than the wavelength of a laser (Rayleigh region), the attractive force of a laser is given as 
\begin{equation}
\vec{F}=\frac{1}{2} \alpha \vec{\nabla}E ^2,
\end{equation}
where $\alpha$ is the polarizability of the object, which is almost proportional to the difference between the square of the refractive index and that of the surrounding medium. When the refracive index of the object is larger than that of the surrounding medium, a focused laser can trap the object in the vicinity of the focal point. We measured the refractive indices of various solutions: $n_{\textrm{dextran}}=1.365$, $n_{\textrm{PEG}}=1.363$, $n_{\textrm{water}}=1.332$, where $n_{\textrm{dextran}}$, $n_{\textrm{PEG}}$, $n_{\textrm{water}}$ are the refractive indices of dextran solution (200 mg/ml), PEG solution (200 mg/ml), and pure water, respectively, i.e., $n_{\textrm{dextran}} \gtrsim n_{\textrm{PEG}} \gg n_{\textrm{water}}$. Consequently, the trapping force is in the order $F_{\textrm{dextran}} \gtrsim F_{\textrm{PEG}} \gg F_{\textrm{water}}$. 
\begin{figure}
\begin{center}
  \includegraphics{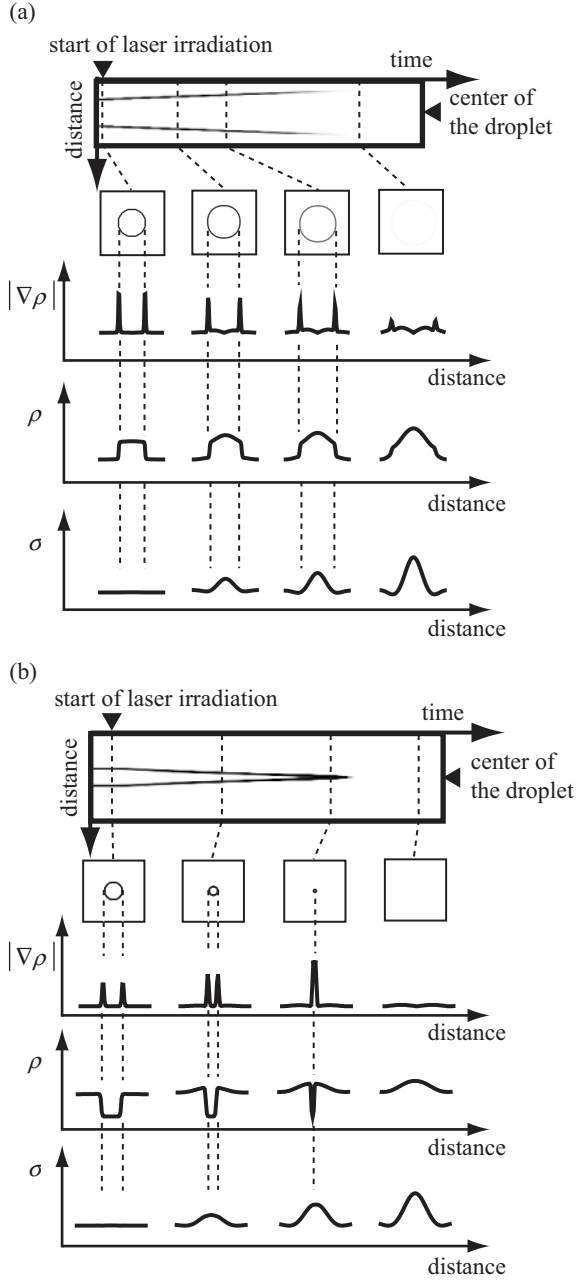}
  \caption{(a) Spatio-temporal image and snapshots of the time development of distribution in the PEG-rich phase and the time development of $|\nabla \rho|$, $\rho$, and $\sigma$, respectively. (b) Spatio-temporal image and snapshots of the time development of the distribution in a dextran-rich phase and the time development of $|\nabla \rho|$, $\rho$, and $\sigma$, respectively. }
  \label{fig:figure4}
\end{center}
\end{figure}%

We can now discuss the two different patterns of droplet disappearance in terms of the Ginzburg-Landau free energy model \cite{safran,onuki} including laser potential. We introduce two order parameters:
\begin{equation}
\rho=(\textrm{[dextran]-[PEG]})/(\textrm{[dextran]+[PEG]}),
\end{equation}
\begin{equation}
\sigma=\textrm{[dextran]+[PEG]} (=1-[\textrm{water}]),
\end{equation}
where [dextran], [PEG] and [water] indicate the volume fraction in the total system of dextran, PEG and water, respectively. Local free energy can be defined in terms of the Ginzburg-Landau model:
\begin{equation}
f_{\rm{local}}(\rho, \sigma)=\rho^4-\beta (\sigma-\sigma_{0})\rho^2,
\label{eq:local_free_energy}
\end{equation}
where $\beta$ is a coupling constant, and $\sigma_{0}$  indicates the threshold value of $\sigma$ , i.e., when $\sigma$ is smaller than $\sigma_{0}$ , $f_{\rm{local}}(\rho, \sigma)$  has only one minimum value, and when $\sigma$  is larger than $\sigma_{0}$ , $f_{\rm{local}}(\rho, \sigma)$  has two local-minimum values, which corresponds to the fact that the system is homogeneous with a lower polymer concentration and undergoes phase separation with a higher polymer concentration. The total free energy including the laser potential can be written as:
\begin{multline}
F_{1}(\rho; \sigma, I(r)) =\int dr\bigl(\rho^4-\beta (\sigma-\sigma_{0})\rho^2\\ -\alpha_{1} \rho I(r) +\frac{1}{2}\gamma_{1} |\nabla \rho|^2 \bigr),
\label{eq:rho_free_energy}
\end{multline}
\begin{multline}
F_{2}(\sigma, I(r))=\int dr \bigl(-\alpha_{2}\sigma I(r)+\frac{1}{2}\gamma_{2}|\nabla \sigma|^2 \bigr),
\label{eq:sigma_free_energy}
\end{multline}
where $F_{1}(\rho; \sigma, I(r))$ is the free energy for $\rho$, and $F_{2}(\sigma, I(r))$ is the free energy for $\sigma$. $\alpha_{1}$, $\alpha_{2}$, $\gamma_{1}$, and $\gamma_{2}$ are constants, and $I(r)=P\exp\left(-r^2/r_{0}^2 \right)$  indicates the laser potential of the Gaussian beam, where $P$, $r_{0}$, and $r$ are laser power, beam waist, and distance from the focus, respectively. $-\alpha_{1}\rho I(r)$ and $-\alpha_{2}\sigma I(r)$ show the effects of laser potential, and indicate that dextran is attracted to the laser field more than PEG, and that polymers are stabilized under laser potential compared to water, respectively.

Due to the relationship among $n_{\textrm{PEG}}$, $n_{\textrm{dextran}}$, and $n_{\textrm{water}}$, the trapping forces exhibit the same relationship as $F_{\textrm{dextran}} \gtrsim F_{\textrm{PEG}} \gg F_{\textrm{water}}$. As a consequence, $\sigma$ changes much faster than $\rho$. Considering the dynamics of $\rho$, the law of conservation cannot be achived for $\rho$  because $\sigma$ changes more rapidly. Therefore, $\rho$ can be treated as a nonconserved parameter, while $\sigma$ is a conserved parameter. Considering all of these points, the evolution equations for $\rho$ and $\sigma$ can be derived from Eqs. (\ref{eq:rho_free_energy}) and (\ref{eq:sigma_free_energy}):
\begin{align}
\frac{\partial \rho}{\partial t}&=-L_{1}\frac{\delta F_{1}}{\delta \rho}\notag  \\
&=-L_{1} (4\rho^3-2\beta(\sigma-\sigma_{0})\rho-\alpha_{1}I(r)-\gamma_{1}\nabla^2 \rho), 
\label{eq:rho_time_development} \\
\frac{\partial \sigma}{\partial t}&=L_{2}\nabla^2\left(\frac{\delta F_{2}}{\delta \sigma} \right) \notag \\
&=L_{2}\nabla^2(-\alpha_{2} I(r)-\gamma_{2}\nabla^2 \sigma),
\label{eq:sigma_time_development}
\end{align}
where $L_{1}$ and $L_{2}$ are the typical time-scales for changes in $\rho$ and $\sigma$, respectively.

Based on the above equations, we performed numerical calculations. The initial value of $\rho$ for the dextran-rich droplet is given as $+\rho_{0}$, while that for the PEG-rich droplet is $-\rho_{0}$, where $\rho_{0}$ is a positive constant. We fixed the parameters as follows: $\beta=1.0$, $\sigma_{0}=0.5$, $\alpha_{1}=0.02$, $\alpha_{2}=0.04$, $L_{1}=0.2$, $L_{2}=0.8$, $\gamma_{1}=0.08$, $\gamma_{2}=1.0$, and $r_{0}=20.0$ for both (a) and (b). $\rho_{\rm droplet}=0.15$, $\rho_{\rm bulk}=-0.15$, and $R=20.0$ for (a), and $\rho_{\rm droplet}=0.15$, $\rho_{\rm bulk}=-0.15$, $R=10.0$ for (b), where $\rho_{\rm droplet}$ and $\rho_{\rm bulk}$ indicate the initial value of $\rho$ inside and outside the droplet respectively, and $R$ shows the radius of the droplet.

Figure \ref{fig:figure4} (a) shows a spatio-temporal image and snapshots of the time development of $|\nabla \rho|$  around the dextran-rich droplet and the profiles of $|\nabla \rho|$, $\rho$, and $\sigma$, respectively. Since the experiments were conducted with a phase-contrast microscope, the interface of the observed droplet roughly corresponds to $|\nabla \rho|$. After laser irradiation begins, $\sigma$ begins to have a non-uniform profile, with larger values around the laser focus and smaller values near the interface of the droplet, which means that polymers aggregate due to laser potential, while water is repelled from the focal point. As a consequence, the profile of $|\nabla \rho|$ at the boundary begins to be smooth.

Figure \ref{fig:figure4} (b) shows a spatio-temporal image and snapshots of the change in $|\nabla \rho|$  around the PEG-rich droplet and profiles of $|\nabla \rho|$, $\rho$, and $\sigma$, respectively. As soon as irradiation began, the PEG-rich droplet began to shrink, since the PEG-rich state becomes metastable. The diameter of the droplet decreased linearly with time. These results well reproduce the experimental results.

In summary, when a focused laser was used to irradiate a mixed polymer system, the interface of the droplets disappeared, mainly due to laser potential. We found two different processes of disappearance. These phenomena were described in terms of the Ginzburg-Landau free energy model, including the laser potential term based on a relational expression of refractive indices of the compounds.

The authors thank Professor Takao Ohta (Kyoto University, Japan) for his kind advice on the mechanism. This work is supported in part by a Grant-in-Aid for 21th century COE "Center for Diversity and Universality in Physics" from the Ministry of Education, Culture, Sports, Science, and Technology of Japan.

\end{document}